\documentclass[12pt]{revtex4}

\usepackage[T1]{fontenc}
\usepackage{isolatin1}
\usepackage{graphicx}
\DeclareGraphicsExtensions{.pdf,.mps,.png,.jpg,.gif,.eps}

\begin{document}

\title{High-angular-precision $\gamma$-ray astronomy and polarimetry}

\author{Denis Bernard}
\affiliation{LLR, Ecole Polytechnique, CNRS/IN2P3, 91128 Palaiseau Cedex, France}

\author{Alain Delbart}
\affiliation{CEA/CE-Saclay, DSM-IRFU, 91191 Gif sur Yvette Cedex, France}
 
\date{\today}

\begin{abstract}
We are developing a concept of a ``thin'' detector as a
high-angular-precision telescope and polarimeter for cosmic $\gamma$-rays
above the pair-creation threshold.
\end{abstract}


\maketitle

Since its launch in 2008, the {\em Fermi} Large Area Telescope (LAT)
has been exploring the $\gamma$-ray emission from cosmic sources in
the range 0.1--300 GeV \cite{Atwood:2009ez}.
The mission has renewed our understanding of pulsars, active galactic
nuclei, globular clusters, gamma-ray bursts,
binary stars, supernova remnants and diffuse $\gamma$-ray sources.
The design life is 5 years and the goal for mission
operations is 10 years.
It is time to consider what could be the next space-based mission
after {\em Fermi}.

\section{A $\gamma$-ray telescope}

The LAT is a detector with an effective area of almost one
square meter above 1 GeV.
While, 
in principle, it can operate down to 20 MeV, its detection
efficiency strongly decreases at low energies after the photon selection,
needed to reject the huge background, is applied.
The drastic degradation of the sensitivity of $e^+e^-$ pair telescopes
 from the GeV down to the MeV, together with the drastic degradation
 of the sensitivity of Compton telescopes above the MeV,
 is a well known issue in X/$\gamma$-ray astronomy, and is referred to as the
 ``sensitivity gap'' \cite{Schoenfelder}.
The way out, as far as pair telescopes are concerned, is to improve
the angular resolution, so as to improve the background rejection for
pointlike sources.

The mapping of the $>$ 70 MeV $\gamma$-ray expected from galactic
 sources from $\pi^0$ decays from high-energy proton collisions with
 matter at rest might solve the mystery of the origin of cosmic rays
 \cite{Weekes:2003gk}.

\section{Polarimetry}

$\gamma$-rays are emitted by cosmic sources in a variety of non-thermal
 processes.
Radiative processes, such as synchrotron radiation or inverse Compton
scattering, provide linearly polarized radiation to some extent, while
nuclear interactions 
end up with non polarized photons.
At lower energies, polarimetry is a key diagnostic in understanding
the properties of a source; e.g., the turbulence of the magnetic field
decreases the observed average polarization fraction. For $\gamma$-rays, 
 this tool is badly missing.
Polarimetry would provide new insight in 
the understanding of a variety of sources such as pulsars 
\cite{Kaspi:2004:IsolNeutronStars,Takata:2007:PolarizationPulsars}, 
$\gamma$-ray bursts (GRBs, e.g.
\cite{Waxman:2003:PolGRBs,Rossi}), 
 AGN 
 \cite{Krawczynski}.

Compton polarimetry is efficient up to a couple of MeV, and
  projects of Compton polarimetry do exist \cite{McConnell},
 but the Compton polarization asymmetry decreases asymptotically as
 the inverse of the energy of the incident photon.

Here the (linear) polarization fraction of the incoming radiation is
obtained from a study of the distribution of the azimuthal angle of
the recoiling electron, in the case of triplet conversions, i.e., $\gamma
e^- \to e^- e^+ e^- $
\cite{Iwata:1993xu,Bernard:2010zz}.

\begin{figure}[htb]
\includegraphics[width=0.456\linewidth]{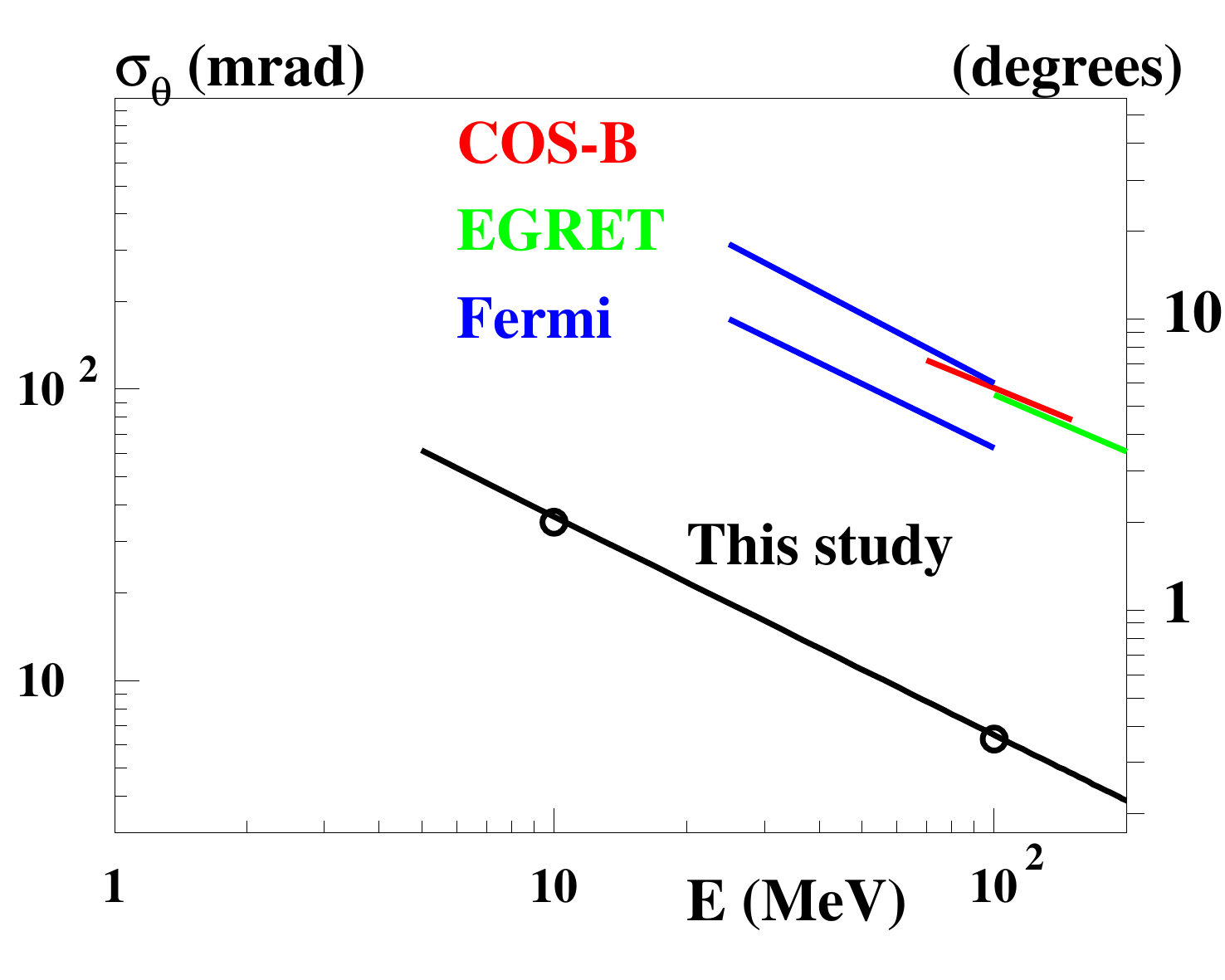}
\hfill
\includegraphics[width=0.456\linewidth]{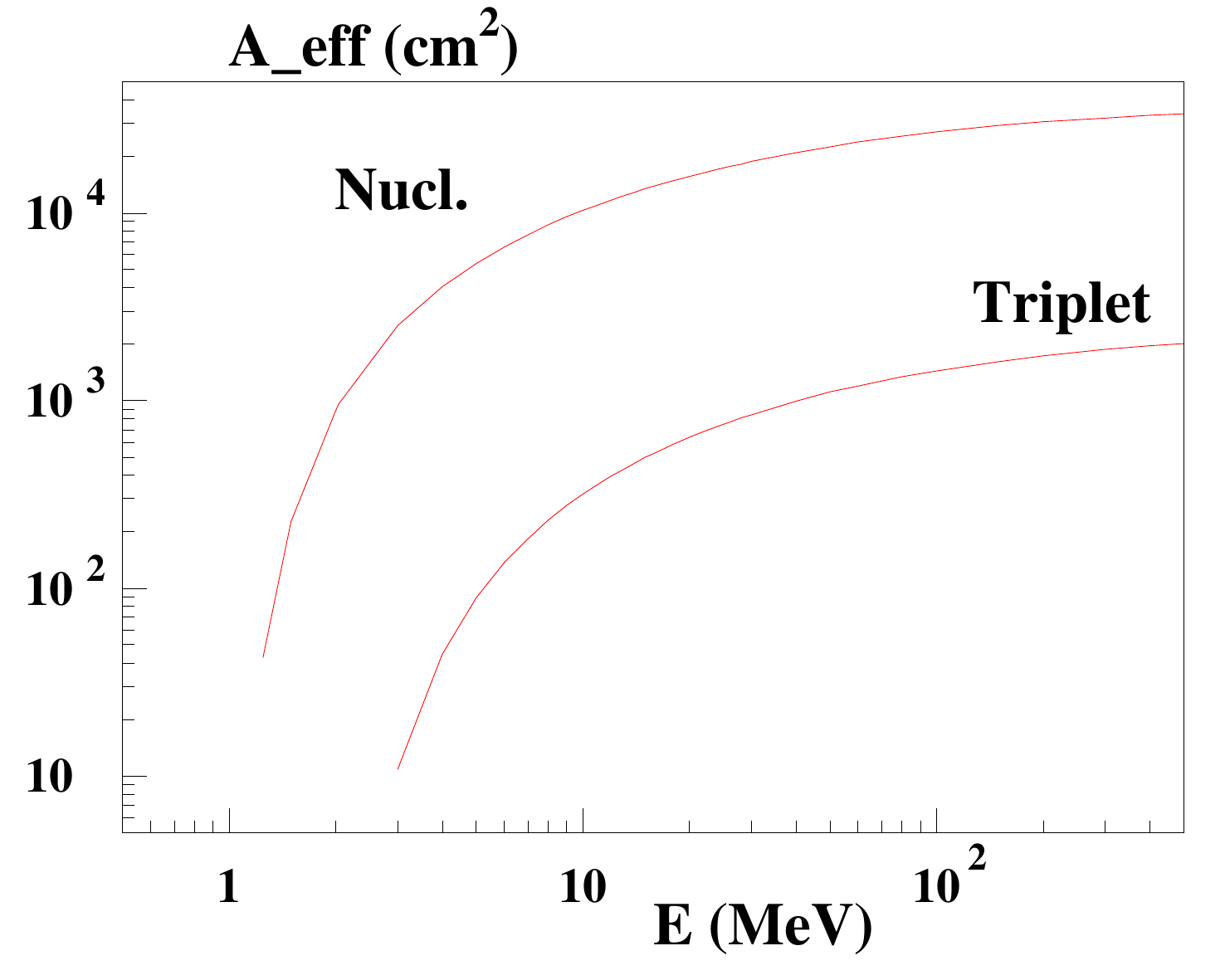}
\caption{\label{fig:resolution:angulaire}Angular resolution as a
function of photon energy compared to past and existing telescopes
(the two {\em Fermi} curves correspond to ``front'' and ``back'' events,
respectively).
The black line is our analytical prediction while the two points are
obtained from simulation (left).
Effective area of a 1 ton Argon-based thin detector, as a function of
 photon energy, for perfect efficiency ($\epsilon = 1$) (right). }
\end{figure}

\section{The detector}

A high-pressure time projection chamber (TPC) i.e., an homogeneous, 3D finely
instrumented detector is a well suited detector.

\begin{itemize}
\item 
The effective area is larger than one square meter per ton over most of
the energy range
(fig. \ref{fig:resolution:angulaire}, right), i.e. larger than that of the LAT.

\item 
The angular resolution is improved by one order of magnitude
(fig. \ref{fig:resolution:angulaire}, left), and therefore the
background rejection factor for point-like sources by two orders of
magnitude.

\item 
A TPC is a deadtime-free GRB detector, as even two simultaneous
incoming photons can be easily detected.

\item 
A TPC is a thin detector\footnote{The interaction length, on the
energy range considered, is smaller than the thickness.} (Compton,
``nuclear'' pair, triplet processed are not in competition with each
other), with effective area simply proportional to the cross section.
It has a high efficiency, a $4\pi$ sr acceptance (reduced to $2\pi$ sr
if operated close to the earth). It is radiation-hard and flux-hard
enough to be used in HEP, more than enough for use in space.

\item 
The energy measurement that is obtained from the multiple measurement
of multiple scattering of the tracks in the detector at low energy
must be complemented by an additionnal system at higher energy, which
will be chosen to be a ``thin'' system too, so as to not kill the mass
budget, i.e., either a magnetic spectrometer, or a
transition-radiation detector (TRD), that can operate up to hundreds
of GeV \cite{Wakely:2004gg}.
\end{itemize}

\begin{figure}[htb]
\begin{center} 
\includegraphics[width=0.7456\linewidth]{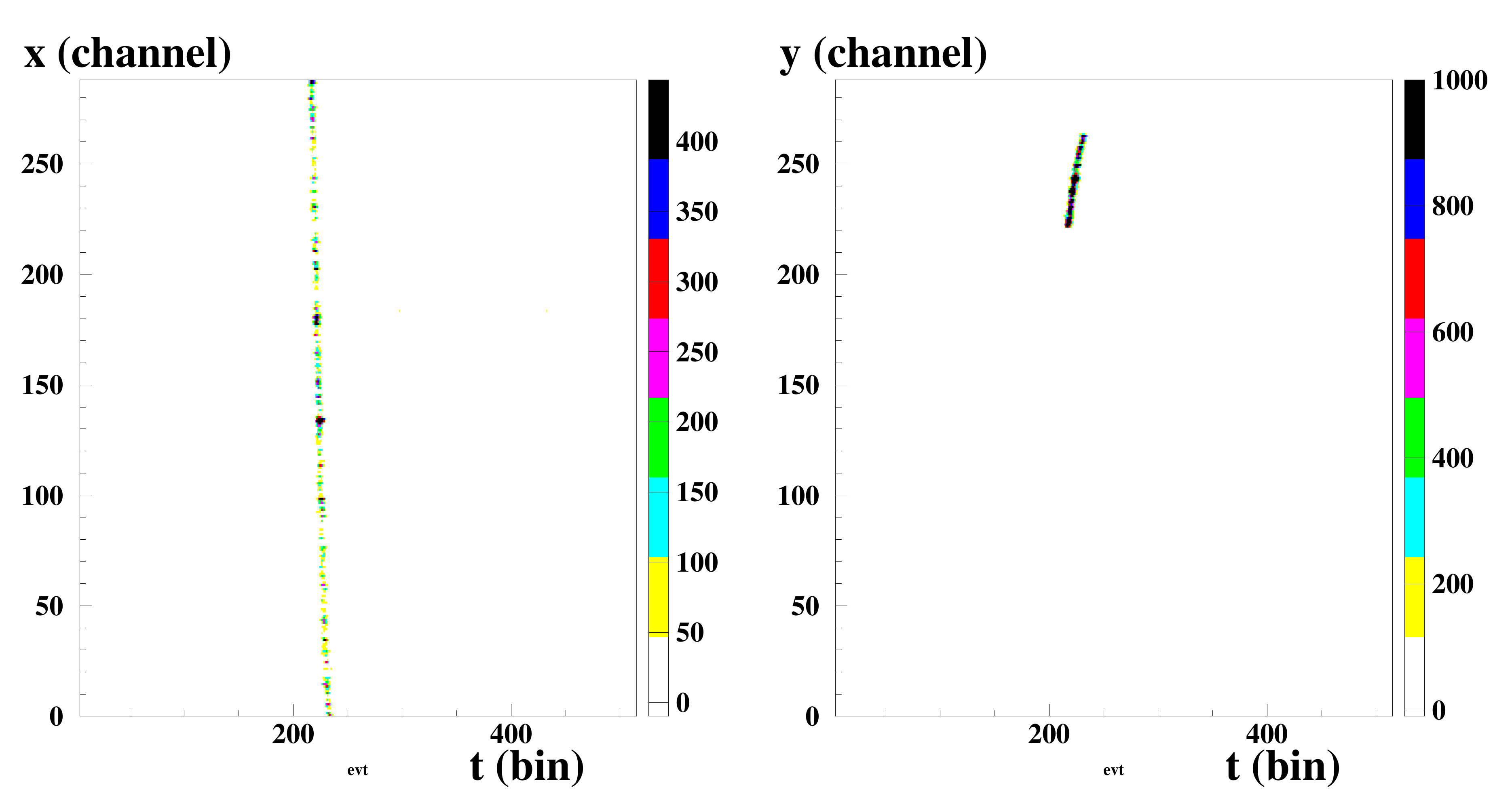}
\caption{\label{fig:onetrack} Cosmic ray crossing the demonstrator.
Left : $(x,t)$. 
Right : $(y,t)$. }
\end{center} 
\end{figure}

\section{The present ``ground'' phase of the project}

A demonstrator is presently being commissionned, consisting of a 5 bar
 Argon-based cubic TPC, with a size of 30 cm, and with a pitch,
 sampling frequency and diffusion-induced resolution of about 1 mm.
The amplification is performed with a ``bulk'' micromegas mesh
 \cite{bulk}, the signal collected by two orthogonal strip sets, and
 digitized with a chip \cite{AFTER} developed for T2K
 \cite{Abgrall:2010hi}.

After we have characterized its performance as a tracker
under (charged) cosmic rays in the laboratory
(Fig. \ref{fig:onetrack})
 we will expose it to a beam of linearly
polarized $\gamma$ rays \cite{NewSUBARU,deJager:2007nf}, aiming at: 
\begin{itemize}
\item validating the technique ($\gamma$ astronomy and polarimetry).
\item obtaining the first measurement of the polarization asymmetries
 in the low energy part of the spectrum (few-MeV -- few 10's MeV),
 where the signal peaks, given the spectra of cosmic sources and where the
 polarization asymmetry is rapidly increasing, and the approximations
 used in the theoretical calculations (screening, Born) are to be
 validated.

\end{itemize}





\bibliographystyle{elsarticle-num}



\end{document}